\documentclass{aastex631}

\usepackage[T1]{fontenc}
\usepackage[figure,figure*]{hypcap}
\usepackage{graphicx}
\usepackage{subfigure}
\usepackage{tabularx}
\usepackage{needspace,enumitem}

\shorttitle{VLA Exoplanets Survey}
\shortauthors{Cendes et al.}
\graphicspath{{./}{figures/}}

\begin{document}

\title{A Pilot Radio Search for Magnetic Activity in Directly Imaged Exoplanets}

\correspondingauthor{Yvette Cendes}
\email{yvette.cendes@cfa.harvard.edu}

\author[0000-0001-7007-6295]{Y. Cendes}
\affiliation{Center for Astrophysics | Harvard \& Smithsonian,
Cambridge, MA 02138, USA}

\author[0000-0003-3734-3587]{P. K. G. Williams}
\affiliation{Center for Astrophysics | Harvard \& Smithsonian,
Cambridge, MA 02138, USA}
\affiliation{American Astronomical Society, 1667 K Street NW, Suite 800
Washington, DC 20006 USA}

\author[0000-0002-9392-9681]{E. Berger}
\affiliation{Center for Astrophysics | Harvard \& Smithsonian,
Cambridge, MA 02138, USA}

\begin{abstract}
We present the first systematic search for GHz frequency radio emission from directly imaged exoplanets using Very Large Array (VLA) observations of sufficient angular resolution to separate the planets from their host stars. We obtained results for five systems and eight exoplanets located at $\lesssim 50$ pc, through new observations (Ross 458, GU Psc, and 51 Eri) and archival data (GJ 504 and HR 8799).  We do not detect radio emission from any of the exoplanets, with $3\sigma$ luminosity upper limits of $(0.9-23)\times10^{21}$ erg s$^{-1}$. These limits are comparable to the level of radio emission detected in several ultracool dwarfs, including T dwarfs, whose masses are only a factor of two times higher than those of the directly-imaged exoplanets.  Despite the lack of detections in this pilot study, we highlight the need for continued GHz frequency radio observations of nearby exoplanets at $\mu$Jy-level sensitivity. 
\end{abstract}


\section{Introduction} 
\label{sec:intro}

Radio observations are a promising means to probe and measure magnetic fields in exoplanets. The properties of these fields are challenging to predict, but their detection could provide essential insight into the internal structure of exoplanets and their habitability (e.g., \citealt{zarka1998,Lazio2019}).  In our own solar system, radio emission at low frequencies ($\sim$MHz) has been observed from all planets that have a magnetic field \citep{Wu1979,zarka1998,t06}, with the emission from Jupiter often considered for scaling relations that can be applied to exoplanets \citep{Lazio2004,Turner2019}.  However, most searches to date have yielded only upper limits \citep[see][and references therein]{Griessmeier2017,Zarka2015}. 
Recently, some possible detections have been reported at low frequencies ($\sim$tens of MHz) with the Low Frequency Array \citep[LOFAR; ][]{Turner2021,Vedantham2020} that could be indicative of exoplanet emission, where the emission mechanism is thought to be similar to solar system analogues. However, the resolution of LOFAR ($\sim45$") is not sufficient to pinpoint precisely where the emission originates from within these candidate systems, and the origin still could be from the host star \citep{Turner2021}.

In contrast, detections of radio emission from brown dwarfs at GHz frequencies has been more fruitful. Starting with the first unexpected detection of the brown dwarf LP944-20 \citep{bbb+01}, GHz frequency observations (primarily with the Very Large Array; VLA) have been used to probe the magnetospheres of very low mass stars and brown dwarfs (collectively, ``ultracool dwarfs'', UCDs; e.g., \citealt{berger2002,brr+05,Berger2006,Berger2010,Kao2018}).  In particular, over the past several years, radio emission has been detected from mid-T dwarfs \citep{rw12,wbz13,rw16,wgb17,Kao2019}, whose effective temperatures and masses are comparable to those of some directly-imaged exoplanets.   Emission at GHz frequencies, similar to that seen from UCDs, has not yet been detected from exoplanets \citep{Bastian2000,Lazio2010,Stroe2012,Griessmeier2017}.


Building on the similarities in physical parameters between detected UCDs and giant exoplanets, and the possibility that UCD magnetospheres are scaled-up analogues of planetary magnetospheres \citep{w17}, we present here the first GHz frequency survey of nearby directly-imaged exoplanets, using observations of sufficient angular resolution to separate the planets from their host stars.  The paper is structured as follows. In \S\ref{sec:obs} we present our target selection and VLA observations, and discuss additional relevant archival data. In \S\ref{sec:results} we present the results of the observations, including subsequent follow-up observations of the 51 Eri system to vet a potential candidate.  In \S\ref{sec:discussion} we discuss our findings in the context of UCD radio emission, and we present our conclusions in \S\ref{sec:conclusions}.

\begin{deluxetable*}{l|C|c|c|c|c|c}
\tablecaption{Exoplanets Targeted in this Work}
\tablecolumns{7}
\tablehead{
\colhead{Name} & 
\colhead{Mass} &
\colhead{Distance} &
\colhead{Stellar Mass} &
\colhead{Spec.~Type}&
\colhead{Semi-Major Axis}&
\colhead{$\Delta\theta$} \\
\colhead{ } &
\colhead{(M$_\mathrm{Jup}$)} &
\colhead{(pc)} &
\colhead{(M$_{\odot}$)} &
\colhead{}&
\colhead{(AU)}&
\colhead{(arcsec)}
}
\startdata
Ross 458 c & 8^{+3}_{-2} & 11.51$\pm$0.02 & 0.56$\pm$0.02& M0.5+M7 & 1168 &102 \\
\hline
GU Psc b & 11.3\pm1.7 & 47.55$\pm$0.16 &0.325$\pm$0.025& M3 & 2000$\pm$200 & 42$\pm$4\\
\hline
51 Eri b & 2.6\pm0.3 & 29.75$\pm$0.12 & 1.75$\pm$0.05&  F0IV & 13.2$\pm$0.2 & 0.46$\pm$0.01\\
\hline
GJ 504 b& 4.0^{+4.5}_{-1.0} & 17.53$\pm$0.08 & 1.22$\pm$0.08 & G0V & 43.5 & 2.48$\pm$0.03\\
\hline
HR 8799 b & 7^{+4}_{-2} & 41.24$\pm$0.15 & $1.61^{+0.32}_{-0.21}$ & A5V & 67.96$\pm$1.85& 1.65$\pm$0.01\\
\hline
HR 8799 c & 10\pm3 & " & " & " & 42.81$\pm$1.16& 0.92$\pm$0.01\\
\hline
HR 8799 d & 10\pm3 & " & " & " & 26.97$\pm$0.73& 0.58$\pm$0.01\\
\hline
HR 8799 e & 10^{+7}_{-4} & " & " & " & 16.99$\pm$0.46& 0.40$\pm$0.01\\ 
\enddata
\tablecomments{Information for Ross 458 from \citet{gmh+10,Rodriguez2011}, for 51 Eri from \citet{mgb+15,Maire2019}, for GJ 504 from \citet{Kuzuhara2013,Zurlo2016}, and for HR 8799b,c,d,e from \citet{Marois2008}.}
\label{tab:planets}
\end{deluxetable*}

\section{Sample Selection and Observations}
\label{sec:obs}

We constructed our target sample using the NASA Exoplanet Archive \footnote{https://exoplanetarchive.ipac.caltech.edu/}, which currently contains 51 exoplanetary systems that are published and confirmed via direct imaging.  We filtered this list further to include systems with a planetary mass consistent with $M_p<13~M_{Jup}$, limited to a distance of $d<50$ pc, and a declination of $\delta>-25^{\circ}$ to enable long observations with the VLA.  These criteria resulted in five systems containing 8 exoplanets.  The properties of these systems are listed in Table~\ref{tab:planets}.

For the previously unobserved systems (Ross 458, GU Psc, and 51 Eri), we obtained VLA observations for 3 hours each in the C-band ($4-8$ GHz; Project 18A-318; PI: Williams).  These observations were obtained in the A configuration, to ensure that the exoplanets can be resolved from their host stars. The details of the observations are summarized in Table~\ref{tab:data}. We processed the data using standard procedures in the Common Astronomy Software Application package (CASA; \citealt{McMullin2007}) accessed through the python-based \texttt{pwkit} package\footnote{https://github.com/pkgw/pwkit} \citep{Williams2017}.  We performed bandpass and flux density calibration using either 3C286 or 3C147 as the primary calibrator for all observations and frequencies. The complex gain calibrators were J1254$+$1141, J0112$+$2244, and J0423$-$0120 for Ross~458, GU~Psc, and 51~Eri, respectively. After applying \textit{a priori} flags, we flagged for radio frequency interference (RFI) using the automated \textsf{aoflagger} tool \citep{odbb+10, ovdgr12}. We imaged the datasets using the CASA imager with multi-frequency synthesis \citep[MFS;][]{the.mfs} and $w$-projection with 128 planes \citep{cgb05}. The cell sizes were $0.08''$ for Ross~458 and $0.06''$ for GU~Psc and 51~Eri. The final images contained only a few sources (Figures~\ref{fig:ross458} and \ref{fig:radio-image}), for which we obtained flux densities and uncertainties using the {\tt imtool fitsrc} command within {\tt pwkit}.  

Several archival JVLA observations exist for GJ 504 and HR 8799; we chose the observations closest to C-band and longest in duration for each target, and analyzed them with {\tt pwkit} as described above.  For HR8799, we identified an S-band ($2-4$ GHz) observation from 2012 November 1 (project code: 12B-188; PI: Ricci)  In the case of GJ 504, we use an S-band observation from 2016 March 3 taken in C configuration with an on-source duration of 0.33 hours (Project 16A-078; PI: Bastian).

\begin{deluxetable*}{l|c|c|c|c|c|c|c|r}
\tablecaption{VLA Observations \label{tab:data}}
\tablecolumns{8}
\tablehead{
\colhead{Name} & 
\colhead{Date} & 
\colhead{Code} &
\colhead{Band} &
\colhead{Config.} &
\colhead{Duration} &
\colhead{$F_\nu$} &
\colhead{$L_\nu$} &
\colhead{$L$} \\
\colhead{} &
\colhead{} &
\colhead{} &
\colhead{} &
\colhead{} &
\colhead{(hr)} &
\colhead{($\mu$Jy)}&
\colhead{(erg s$^{-1}$ Hz$^{-1}$)}&
\colhead{(erg s$^{-1}$)}
}
\startdata
Ross 458 & Mar-26-2018 & 18A-318 & C & A & 3.0 & $<9$ & $<1.4\times10^{12}$ & $<8.6\times10^{21}$ \\
\hline
GU Psc &Mar-03-2018 & 18A-318 & C & A & 3.0 & $<6$ & $<1.6\times10^{13}$ & $<9.7\times10^{22}$ \\
\hline
51 Eri & Mar-02-2018 & 18A-318 & C & A & 3.0 & $<6$ & $<6.4\times10^{12}$ & $<3.8\times10^{22}$ \\
\hline
GJ 504 & Mar-03-2016 & 16A-078 & S & C & 0.33 & $<210$ & $<7.7\times10^{13}$ & $<2.3\times10^{23}$ \\
\hline
HR 8799 & Nov-01-2012 & 12B-188 & S & A & 2.76 & $<24$ & $<8.7\times10^{12}$ & $<2.6\times10^{22}$ \\
\hline
\enddata
\tablecomments{Upper limits are 3$\sigma$.  
}
\end{deluxetable*}

\section{Results}
\label{sec:results}

\begin{figure}[t]
\centering
  \includegraphics[width=1\textwidth]{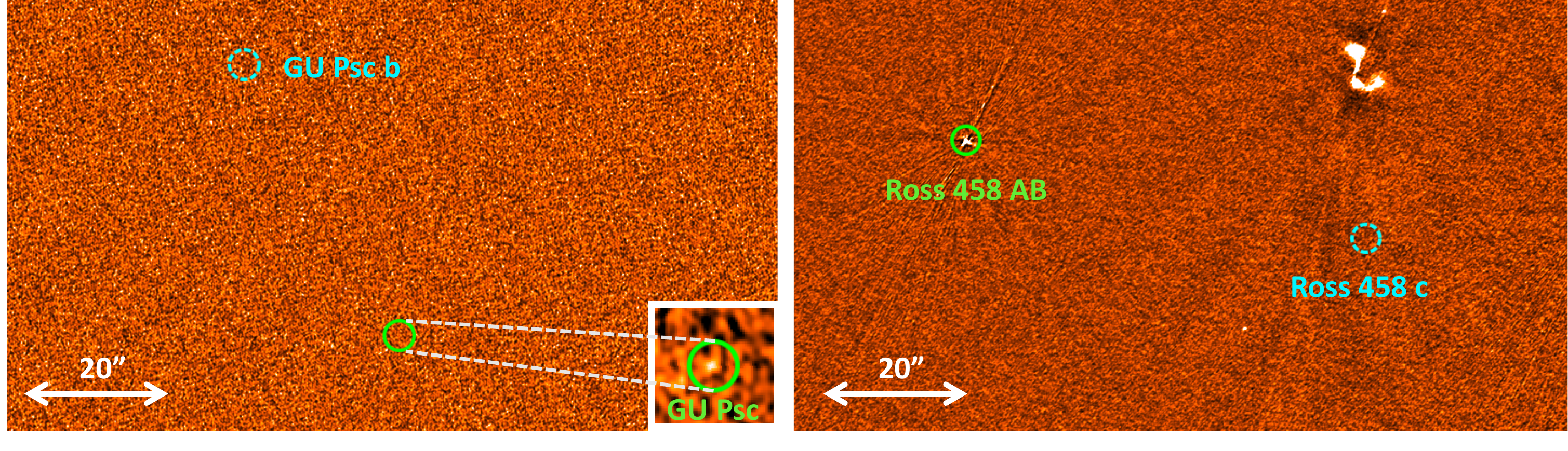}
  \caption{VLA C-band images of the GU Psc ({\it Left}) and Ross 458 ({\it Right}) systems, spanning $130\times75$'' and oriented with north up and east to the left.  We show the position of the star (green circle) and the position of the exoplanet (blue circle), and provide a zoomed-in insert of the GU Psc source on left because it is not visible at the scale of the larger image.  The positional data for Ross 458 are from \citet{Goldman2010} and for GU Psc from \citet{Naud2014}.
  }
  \label{fig:ross458}
\end{figure}

In all 5 systems (8 exoplanets), there is no evidence for radio emission at the planet positions. The results are summarized in Table~\ref{tab:data}. For GU Psc, we detect weak emission ($11\pm 2$ $\mu$Jy) from the host star, and place a $3\sigma$ limit of $\lesssim 6$ $\mu$Jy at the exoplanet position; see right panel of Figure~\ref{fig:ross458}.  In the Ross 458 system, we detect bright emission ($1.37\pm 0.03$ mJy), coincident with the host star position, and place a limit of $\lesssim 9$ $\mu$Jy at the exoplanet position (left panel of Figure~\ref{fig:ross458}).  Ross 458 AB is a M0.5+M7 binary with a separation of $\approx 0.48''$ \citep{Beuzit2004}, where the primary is a known flare star \citep{White1989}.

For HR 8799, we obtained a $3\sigma$ upper limit of $\lesssim 24$ $\mu$Jy.  As the background noise was consistent across the entire image, we adopt a single upper limit on the luminosity for all planets in the HR 8799 system.  For GJ 504, we find a $3\sigma$ upper limit of $\lesssim 210$ $\mu$Jy, due to the short duration of the observation combined with RFI.  

\begin{figure}[t]
\centering
  \includegraphics[width=0.8\textwidth]{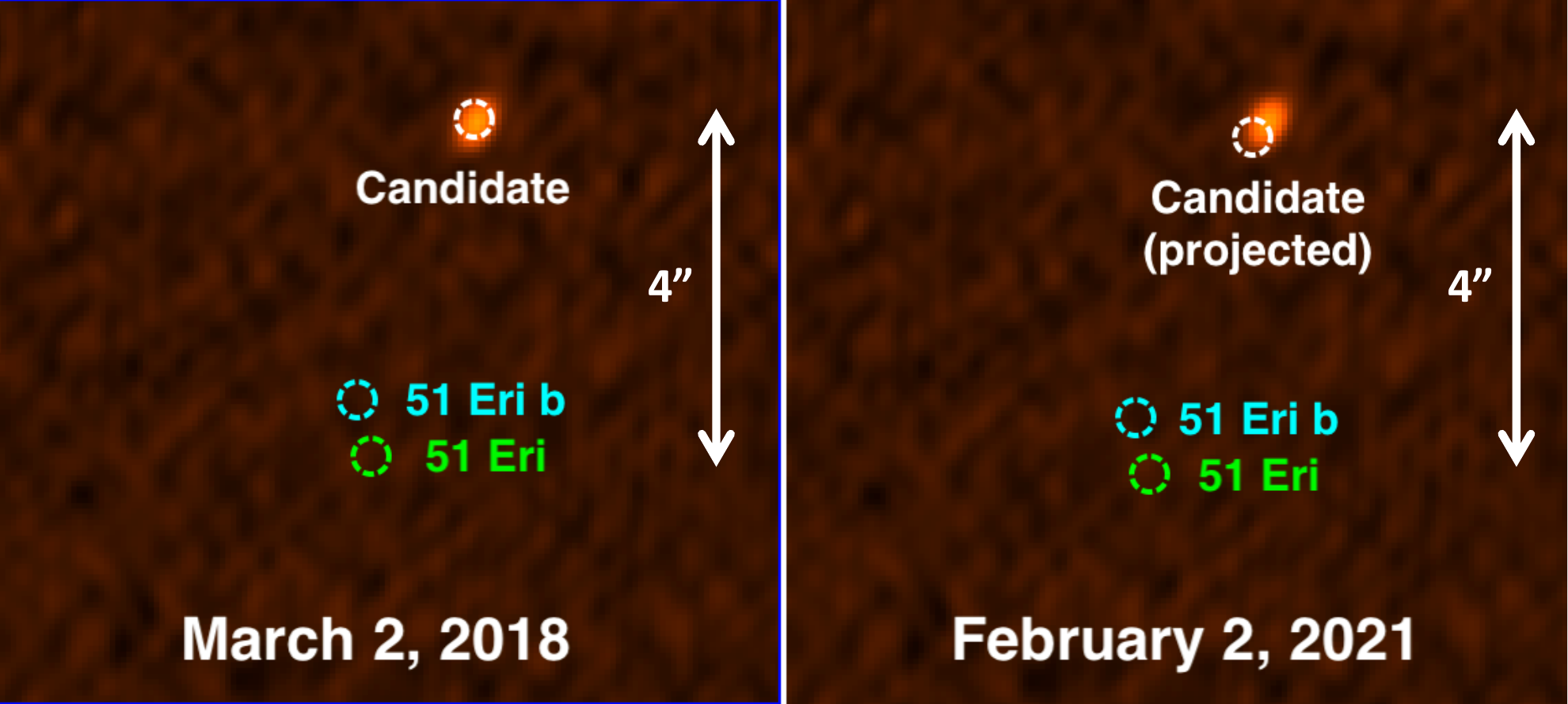}
  \caption{VLA C-band images of the 51 Eri system from 2018 March 2 ({\it Left}) and 2021 February 2 ({\it Right}) spanning $8''$ on a side.  Also shown are the positions of 51 Eri (green circle) and its known directly-imaged exoplanet 51 Eri b (blue circle) based on positional information from \citet{Maire2019}. We detect a radio source in the field (white circle, left), at a location that is outside of the field of the high-contrast direct-imaging instruments.  The follow-up observations indicate that this source did not move as expected if it was associated with the 51 Eri system (white circle, right). 
  }
  \label{fig:radio-image}
\end{figure}

In the case of 51 Eri, we did not detect radio emission at the position of the star or exoplanet, with a limit of $\lesssim 6$ $\mu$Jy (Figure~\ref{fig:radio-image}).  However, we did detect a radio source with a flux density of $89\pm 3$ $\mu$Jy about $4.8''$ away from 51 Eri (Figure~\ref{fig:radio-image}), which varied by a factor of about 2 during the course of our 3 hr observation.  The location of this source is outside of the field of view of the original high-resolution exoplanet discovery image. The source is also sufficiently close to the 5.26 mag host star that we cannot clearly identify a counterpart in archival {\it Hubble Space Telescope} images obtained with ACS/HRC using the Coron1.8 aperture and F606W filter (Program 10487; PI: Ardila).

To determine if this source is an unrelated background source (a scintillating active galactic nucleus (AGN) given its variability) or associated with the 51 Eri system (potentially a more distant exoplanet), we carried out follow-up VLA observations on 2021 February 2, $\approx 2.9$ yr after the first observation (Project 20B-249; PI: Cendes). Using Gaia DR2, we find that the proper motion of 51 Eri is $\delta{\rm RA}=44.35\pm 0.23$ mas yr$^{-1}$ and $\delta{\rm Dec}=-63.83\pm 0.18$ mas yr$^{-1}$ \citep{Bailer-Jones2018}, or a total motion of about 227 mas over the time baseline of our observations. We note that 51 Eri b moved $\sim 15^{\circ}$ in its orbit during this period \citep{Maire2019}, which is negligible compared to the overall proper motion of the system. We detect the same nearby source in the second observation (Figure~\ref{fig:radio-image}) at a position fully consistent with the first observation, and place a $3\sigma$ upper limit on its motion of $\lesssim 20$ mas, an order of magnitude smaller than if it were associated with the 51 Eri system. We thus conclude that this source is most likely a background AGN.

\section{Discussion}
\label{sec:discussion}

\begin{figure}[t]
\centering
  \includegraphics[width=0.9\textwidth]{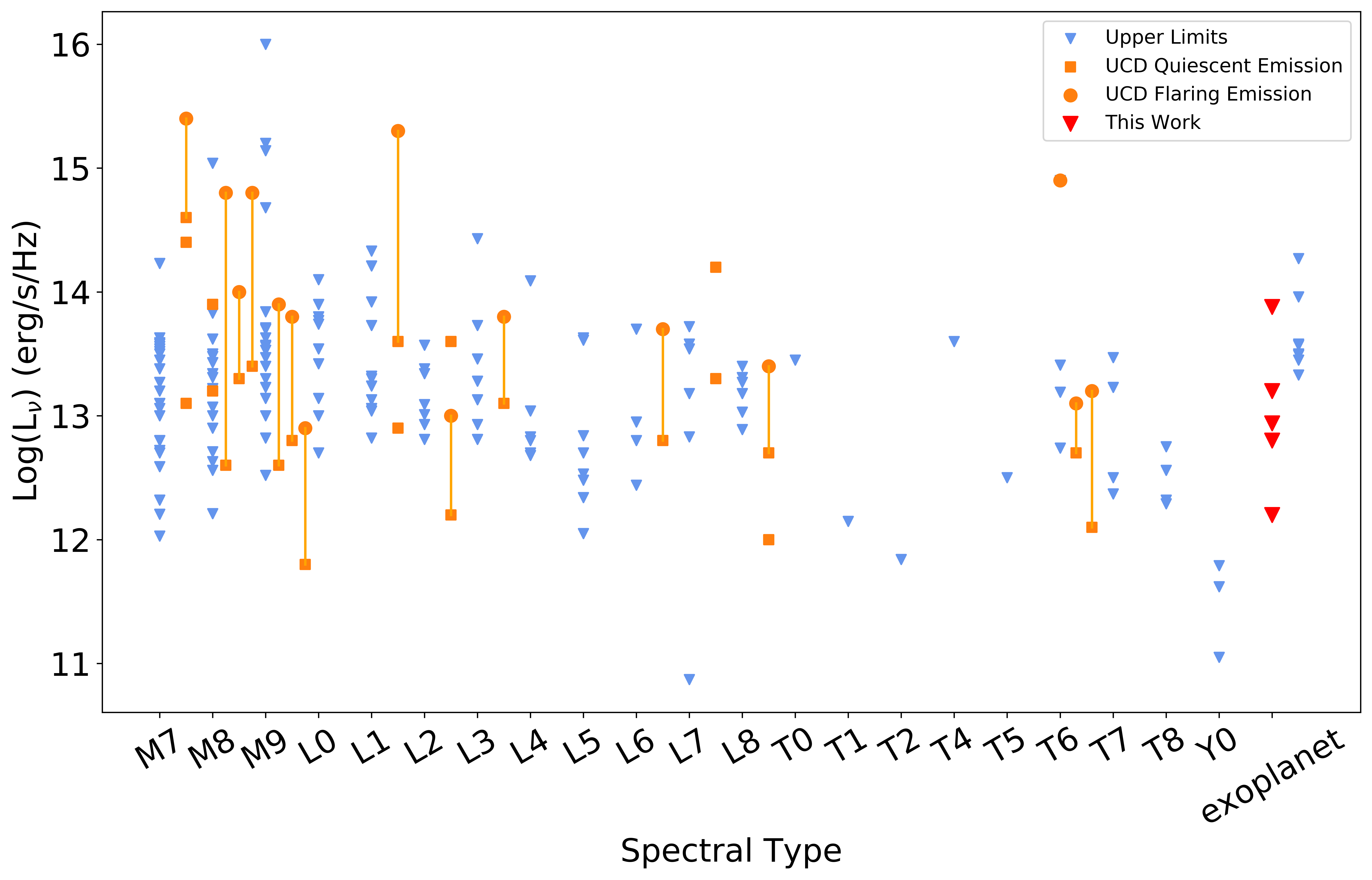}
  \caption{Radio spectral luminosity upper limits for the exoplanets in our survey (red triangles) compared to observations of UCDs plotted against spectral type (circles: flares; squares: quiescent; triangles: upper limits; lines connect flaring and quiescent emission from a single UCD), and constraining upper limits for exoplanets at GHz frequencies. The detected UCDs have been slightly shifted in spectral type for clarity.  Data for UCDs are from \citet{Pineda2017}, \citet{Kao2019} and references therein, and for the exoplanet upper limits from \citet{Bastian2000,Stroe2012,Lazio2010}.}
  \label{fig:exo-lumin}
\end{figure}

Radio emission at GHz frequencies has been detected from several UCDs spanning late-M, L, and T spectral types; see Figure~\ref{fig:exo-lumin}.  The observed radio emission includes both quasi-steady quiescent emission (sometimes varying sinusoidally with the object's rotation) and flaring emission (e.g., \citealt{bbb+01,berger2002,Berger2006,Berger2010,McLean2012,Pineda2017}), with most, but not all, UCDs exhibiting both.  The flaring emission itself includes both stochastic and periodic flares, with a range of durations and circular polarization fractions  (e.g., \citealt{berger2002,hbl2007,brp+09,rw2013}). The overall detection fraction of UCDs is $\sim 5-10\%$, and potentially depends on spectral type although this is based on a small number of objects (e.g., \citealt{berger2002,Berger2006,McLean2012,wcb2014,Pineda2017}).  There also appears to be a correlation between the presence of radio emission and rapid rotation \citep{McLean2012}. 

\begin{figure}[t]
\centering
  \includegraphics[width=0.55\textwidth]{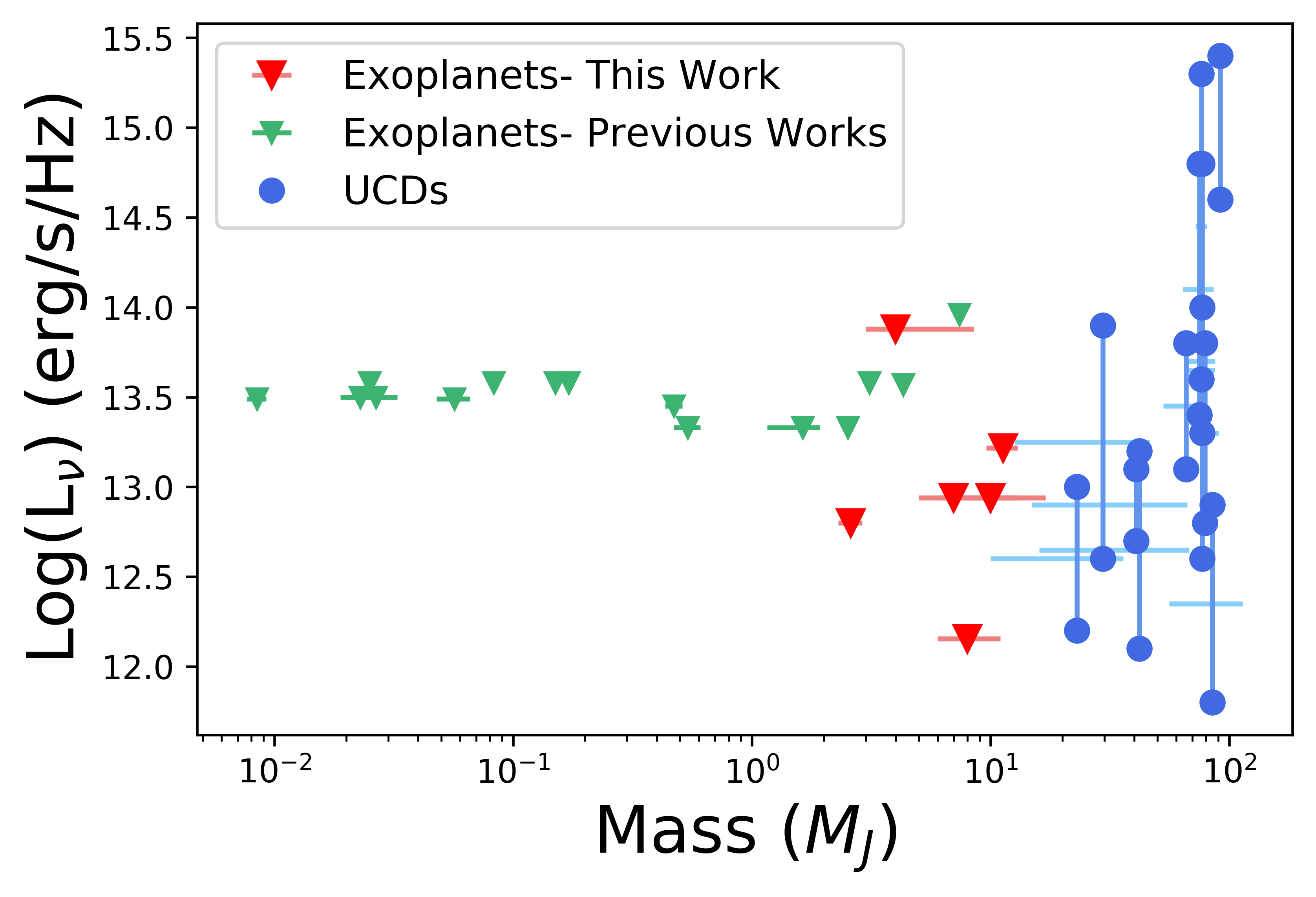}
  \caption{Radio spectral luminosity as a function of mass for the exoplanets in our survey (red triangles), for detected UCDs (blue circles; flares and quiescent emission), and for previously-observed exoplanets (green triangles).  Data for the UCDs is from \citet{Pineda2017} and references therein, while the data for the other exoplanet are from \citet{Bastian2000,Pineda2017,Lazio2010}.  The exoplanet masses for our sample are from the references listed in Table~\ref{tab:planets}, while the masses for the other exoplanets are from \citet{Gregory2010,Borsa2015,Stassun2017,Bourrier2018,Luhn2019,Butler2006,Rosenthal2021,Feng2020}.}
  \label{fig:exo-mass}
\end{figure}

In Figure~\ref{fig:exo-lumin} we plot the quiescent and flaring radio luminosities as a function of spectral type for all UCDs observed at GHz frequencies to date.  The radio spectral luminosities span $\approx 10^{12}-10^{15.5}$ erg s$^{-1}$ Hz$^{-1}$, including for the 3 detected T dwarfs.  The spectral luminosity upper limits for our sample of directly-imaged exoplanets span $\approx 10^{12}$ to $10^{14}$ erg s$^{-1}$ Hz$^{-1}$.  For our best limit (Ross 458 c) we can rule out emission similar to that of detected UCDs, while for the other systems we cannot rule out the presence of weaker emission that is seen in a few UCDs. In Figure~\ref{fig:exo-mass} we plot the radio spectral luminosities of the detected UCDs (quiescent and flaring), the upper limits for our directly-imaged exoplanet targets, and upper limits for a few other nearby exoplanets that were discovered via the radial velocity or transit methods \citep{Bastian2000,Stroe2012} as a function of mass.  Here we see possible evidence for a systematic decrease in radio luminosity with declining mass, as previously suggested in \citet{Route2016}.  Our upper limits for the exoplanets are generally in agreement with this trend.

Overall, given the observed detection fraction in UCDs, it may not be surprising that none of our targets have been detected.  Moreover, since we do not know the rotation periods of the directly-imaged exoplanets, it is possible that their detection rates could be suppressed by slow rotation. Depending on the nature of the emission, beaming might also be a factor, but this is difficult to evaluate in the absence of detections.

\section{Conclusions}
\label{sec:conclusions}

We report results from the first VLA radio survey of directly-imaged exoplanets at GHz frequencies. None of the 8 exoplanets in our study are detected, with luminosity upper limits of $\lesssim (0.9-23)\times10^{22}$ erg s$^{-1}$.  These upper limits are comparable to the radio luminosities of previously detected UCDs, including T dwarfs, which are comparable in mass and effective temperature to the exoplanets observed here.

If we consider these exoplanets in the context of the typical detection rate of radio emission from volume-limited searches of UCD emission \citep[$5-10\%$; ][]{Route2016}, we find 22 exoplanetary systems would need to be observed to have a $>90\%$ chance of achieving a detection.  We thus conclude that additional observations are imperative to determine whether GHz radio emission extends from UCDs to the exoplanet regime, given that GHz emission from UCDs with the same effective temperatures as exoplanets has been seen, and was unexpected when first found \citep{bbb+01}.  In the southern hemisphere, new radio interferometers with high sensitivity, such as MeerKAT, ASKAP, and eventually SKA, will be able to observe directly-imaged exoplanets that are inaccessible to the VLA.  Additionally, new discoveries via direct imaging and other exoplanet detection methods will allow the sample to be expanded.  For example, for a system at $\lesssim 20$ pc, to achieve a separation of $\gtrsim 0.2''$ from the host star requires a semi-major axis of $\gtrsim 4$ AU.  Several known exoplanets currently match these parameters, and are promising targets for radio observations that can distinguish emission from the planet and from the host star.

\begin{acknowledgements}
The Berger Time-Domain Group at Harvard is supported by NSF and NASA grants.  The National Radio Astronomy Observatory is a facility of the National Science Foundation operated under cooperative agreement by Associated Universities, Inc.  This research has made use of the NASA Exoplanet Archive, which is operated by the California Institute of Technology, under contract with the National Aeronautics and Space Administration under the Exoplanet Exploration Program.
\end{acknowledgements}

\software{CASA (McMullin et al. 2007), pwkit (Williams et al. 2017)}

\bibliography{sample631}{}

\begin{thebibliography}{}
\expandafter\ifx\csname natexlab\endcsname\relax\def\natexlab#1{#1}\fi
\providecommand{\url}[1]{\href{#1}{#1}}
\providecommand{\dodoi}[1]{doi:~\href{http://doi.org/#1}{\nolinkurl{#1}}}
\providecommand{\doeprint}[1]{\href{http://ascl.net/#1}{\nolinkurl{http://ascl.net/#1}}}
\providecommand{\doarXiv}[1]{\href{https://arxiv.org/abs/#1}{\nolinkurl{https://arxiv.org/abs/#1}}}

\bibitem[{{Bailer-Jones} {et~al.}(2018){Bailer-Jones}, {Rybizki}, {Fouesneau},
  {Mantelet}, \& {Andrae}}]{Bailer-Jones2018}
{Bailer-Jones}, C.~A.~L., {Rybizki}, J., {Fouesneau}, M., {Mantelet}, G., \&
  {Andrae}, R. 2018, \aj, 156, 58, \dodoi{10.3847/1538-3881/aacb21}

\bibitem[{{Bastian} {et~al.}(2000){Bastian}, {Dulk}, \&
  {Leblanc}}]{Bastian2000}
{Bastian}, T.~S., {Dulk}, G.~A., \& {Leblanc}, Y. 2000, \apj, 545, 1058,
  \dodoi{10.1086/317864}

\bibitem[{{Berger}(2002)}]{berger2002}
{Berger}, E. 2002, \apj, 572, 503, \dodoi{10.1086/340301}

\bibitem[{{Berger}(2006)}]{Berger2006}
---. 2006, \apj, 648, 629, \dodoi{10.1086/505787}

\bibitem[{{Berger} {et~al.}(2001){Berger}, {Ball}, {Becker}, {Clarke}, {Frail},
  {Fukuda}, {Hoffman}, {Mellon}, {Momjian}, {Murphy}, {Teng}, {Woodruff},
  {Zauderer}, \& {Zavala}}]{bbb+01}
{Berger}, E., {Ball}, S., {Becker}, K.~M., {et~al.} 2001, Natur, 410, 338,
  \dodoi{10.1038/35066514}

\bibitem[{{Berger} {et~al.}(2005){Berger}, {Rutledge}, {Reid}, {Bildsten},
  {Gizis}, {Liebert}, {Mart\'{\i}n}, {Basri}, {Jayawardhana}, {Brandeker},
  {Fleming}, {Johns-Krull}, {Giampapa}, {Hawley}, \& {Schmitt}}]{brr+05}
{Berger}, E., {Rutledge}, R.~E., {Reid}, I.~N., {et~al.} 2005, ApJ, 627, 960,
  \dodoi{10.1086/430343}

\bibitem[{{Berger} {et~al.}(2009){Berger}, {Rutledge}, {Phan-Bao}, {Basri},
  {Giampapa}, {Gizis}, {Liebert}, {Mart{\'\i}n}, \& {Fleming}}]{brp+09}
{Berger}, E., {Rutledge}, R.~E., {Phan-Bao}, N., {et~al.} 2009, \apj, 695, 310,
  \dodoi{10.1088/0004-637X/695/1/310}

\bibitem[{{Berger} {et~al.}(2010){Berger}, {Basri}, {Fleming}, {Giampapa},
  {Gizis}, {Liebert}, {Mart{\'\i}n}, {Phan-Bao}, \& {Rutledge}}]{Berger2010}
{Berger}, E., {Basri}, G., {Fleming}, T.~A., {et~al.} 2010, \apj, 709, 332,
  \dodoi{10.1088/0004-637X/709/1/332}

\bibitem[{{Beuzit} {et~al.}(2004){Beuzit}, {S{\'e}gransan}, {Forveille},
  {Udry}, {Delfosse}, {Mayor}, {Perrier}, {Hainaut}, {Roddier}, {Roddier}, \&
  {Mart{\'\i}n}}]{Beuzit2004}
{Beuzit}, J.~L., {S{\'e}gransan}, D., {Forveille}, T., {et~al.} 2004, \aap,
  425, 997, \dodoi{10.1051/0004-6361:20048006}

\bibitem[{{Borsa} {et~al.}(2015){Borsa}, {Scandariato}, {Rainer}, {Bignamini},
  {Maggio}, {Poretti}, {Lanza}, {Di Mauro}, {Benatti}, {Biazzo}, {Bonomo},
  {Damasso}, {Esposito}, {Gratton}, {Affer}, {Barbieri}, {Boccato}, {Claudi},
  {Cosentino}, {Covino}, {Desidera}, {Fiorenzano}, {Gandolfi}, {Harutyunyan},
  {Maldonado}, {Micela}, {Molaro}, {Molinari}, {Pagano}, {Pillitteri},
  {Piotto}, {Shkolnik}, {Silvotti}, {Smareglia}, {Southworth}, {Sozzetti}, \&
  {Stelzer}}]{Borsa2015}
{Borsa}, F., {Scandariato}, G., {Rainer}, M., {et~al.} 2015, \aap, 578, A64,
  \dodoi{10.1051/0004-6361/201525741}

\bibitem[{{Bourrier} {et~al.}(2018){Bourrier}, {Dumusque}, {Dorn}, {Henry},
  {Astudillo-Defru}, {Rey}, {Benneke}, {H{\'e}brard}, {Lovis}, {Demory},
  {Moutou}, \& {Ehrenreich}}]{Bourrier2018}
{Bourrier}, V., {Dumusque}, X., {Dorn}, C., {et~al.} 2018, \aap, 619, A1,
  \dodoi{10.1051/0004-6361/201833154}

\bibitem[{{Butler} {et~al.}(2006){Butler}, {Wright}, {Marcy}, {Fischer},
  {Vogt}, {Tinney}, {Jones}, {Carter}, {Johnson}, {McCarthy}, \&
  {Penny}}]{Butler2006}
{Butler}, R.~P., {Wright}, J.~T., {Marcy}, G.~W., {et~al.} 2006, \apj, 646,
  505, \dodoi{10.1086/504701}

\bibitem[{{Cornwell} {et~al.}(2005){Cornwell}, {Golap}, \& {Bhatnagar}}]{cgb05}
{Cornwell}, T.~J., {Golap}, K., \& {Bhatnagar}, S. 2005, in Astronomical
  Society of the Pacific Conference Series, Vol. 347, Astronomical Data
  Analysis Software and Systems XIV, ed. P.~{Shopbell}, M.~{Britton}, \&
  R.~{Ebert}, 86+.
\newblock \url{http://adsabs.harvard.edu/abs/2005ASPC..347...86C}

\bibitem[{{Feng} {et~al.}(2020){Feng}, {Butler}, {Shectman}, {Crane}, {Vogt},
  {Chambers}, {Jones}, {Xuesong Wang}, {Teske}, {Burt}, {D{\'\i}az}, \&
  {Thompson}}]{Feng2020}
{Feng}, F., {Butler}, R.~P., {Shectman}, S.~A., {et~al.} 2020, \apjs, 246, 11,
  \dodoi{10.3847/1538-4365/ab5e7c}

\bibitem[{{Goldman} {et~al.}(2010{\natexlab{a}}){Goldman}, {Marsat}, {Henning},
  {Clemens}, \& {Greiner}}]{gmh+10}
{Goldman}, B., {Marsat}, S., {Henning}, T., {Clemens}, C., \& {Greiner}, J.
  2010{\natexlab{a}}, MNRAS, 405, 1140+,
  \dodoi{10.1111/j.1365-2966.2010.16524.x}

\bibitem[{{Goldman} {et~al.}(2010{\natexlab{b}}){Goldman}, {Marsat}, {Henning},
  {Clemens}, \& {Greiner}}]{Goldman2010}
---. 2010{\natexlab{b}}, \mnras, 405, 1140,
  \dodoi{10.1111/j.1365-2966.2010.16524.x}

\bibitem[{{Gregory} \& {Fischer}(2010)}]{Gregory2010}
{Gregory}, P.~C., \& {Fischer}, D.~A. 2010, \mnras, 403, 731,
  \dodoi{10.1111/j.1365-2966.2009.16233.x}

\bibitem[{{Griessmeier}(2017)}]{Griessmeier2017}
{Griessmeier}, J.~M. 2017, in Planetary Radio Emissions VIII, ed. G.~{Fischer},
  G.~{Mann}, M.~{Panchenko}, \& P.~{Zarka}, 285--299, \dodoi{10.1553/PRE8s285}

\bibitem[{{Hallinan} {et~al.}(2007){Hallinan}, {Bourke}, {Lane}, {Antonova},
  {Zavala}, {Brisken}, {Boyle}, {Vrba}, {Doyle}, \& {Golden}}]{hbl2007}
{Hallinan}, G., {Bourke}, S., {Lane}, C., {et~al.} 2007, \apjl, 663, L25,
  \dodoi{10.1086/519790}

\bibitem[{{Kao} {et~al.}(2019){Kao}, {Hallinan}, \& {Pineda}}]{Kao2019}
{Kao}, M.~M., {Hallinan}, G., \& {Pineda}, J.~S. 2019, \mnras, 487, 1994,
  \dodoi{10.1093/mnras/stz1372}

\bibitem[{{Kao} {et~al.}(2018){Kao}, {Hallinan}, {Pineda}, {Stevenson}, \&
  {Burgasser}}]{Kao2018}
{Kao}, M.~M., {Hallinan}, G., {Pineda}, J.~S., {Stevenson}, D., \& {Burgasser},
  A. 2018, \apjs, 237, 25, \dodoi{10.3847/1538-4365/aac2d5}

\bibitem[{{Kuzuhara} {et~al.}(2013){Kuzuhara}, {Tamura}, {Kudo}, {Janson},
  {Kandori}, {Brandt}, {Thalmann}, {Spiegel}, {Biller}, {Carson}, {Hori},
  {Suzuki}, {Burrows}, {Henning}, {Turner}, {McElwain}, {Moro-Mart{\'\i}n},
  {Suenaga}, {Takahashi}, {Kwon}, {Lucas}, {Abe}, {Brandner}, {Egner}, {Feldt},
  {Fujiwara}, {Goto}, {Grady}, {Guyon}, {Hashimoto}, {Hayano}, {Hayashi},
  {Hayashi}, {Hodapp}, {Ishii}, {Iye}, {Knapp}, {Matsuo}, {Mayama}, {Miyama},
  {Morino}, {Nishikawa}, {Nishimura}, {Kotani}, {Kusakabe}, {Pyo}, {Serabyn},
  {Suto}, {Takami}, {Takato}, {Terada}, {Tomono}, {Watanabe}, {Wisniewski},
  {Yamada}, {Takami}, \& {Usuda}}]{Kuzuhara2013}
{Kuzuhara}, M., {Tamura}, M., {Kudo}, T., {et~al.} 2013, \apj, 774, 11,
  \dodoi{10.1088/0004-637X/774/1/11}

\bibitem[{{Lazio} {et~al.}(2019){Lazio}, {Hallinan}, {Airapetian}, {Brain},
  {Clarke}, {Dolch}, {Dong}, {Driscoll}, {Fares}, {Griessmeier}, {Farrell},
  {Kasper}, {Murphy}, {Rogers}, {Shkolnik}, {Stanley}, {Strugarek}, {Turner},
  {Wolszczan}, {Zarka}, {Knapp}, {Lynch}, \& {Turner}}]{Lazio2019}
{Lazio}, J., {Hallinan}, G., {Airapetian}, A., {et~al.} 2019, \baas, 51, 135.
\newblock \doarXiv{1803.06487}

\bibitem[{{Lazio} {et~al.}(2010){Lazio}, {Shankland}, {Farrell}, \&
  {Blank}}]{Lazio2010}
{Lazio}, T. J.~W., {Shankland}, P.~D., {Farrell}, W.~M., \& {Blank}, D.~L.
  2010, \aj, 140, 1929, \dodoi{10.1088/0004-6256/140/6/1929}

\bibitem[{{Lazio} {et~al.}(2004){Lazio}, {Farrell}, {Dietrick}, {Greenlees},
  {Hogan}, {Jones}, \& {Hennig}}]{Lazio2004}
{Lazio}, T.~Joseph, W., {Farrell}, W.~M., {Dietrick}, J., {et~al.} 2004, \apj,
  612, 511, \dodoi{10.1086/422449}

\bibitem[{{Luhn} {et~al.}(2019){Luhn}, {Bastien}, {Wright}, {Johnson},
  {Howard}, \& {Isaacson}}]{Luhn2019}
{Luhn}, J.~K., {Bastien}, F.~A., {Wright}, J.~T., {et~al.} 2019, \aj, 157, 149,
  \dodoi{10.3847/1538-3881/aaf5d0}

\bibitem[{{Macintosh} {et~al.}(2015){Macintosh}, {Graham}, {Barman}, {De Rosa},
  {Konopacky}, {Marley}, {Marois}, {Nielsen}, {Pueyo}, {Rajan}, {Rameau},
  {Saumon}, {Wang}, {Patience}, {Ammons}, {Arriaga}, {Artigau}, {Beckwith},
  {Brewster}, {Bruzzone}, {Bulger}, {Burningham}, {Burrows}, {Chen}, {Chiang},
  {Chilcote}, {Dawson}, {Dong}, {Doyon}, {Draper}, {Duch\^{e} ne}, {Esposito},
  {Fabrycky}, {Fitzgerald}, {Follette}, {Fortney}, {Gerard}, {Goodsell},
  {Greenbaum}, {Hibon}, {Hinkley}, {Cotten}, {Hung}, {Ingraham},
  {Johnson-Groh}, {Kalas}, {Lafreniere}, {Larkin}, {Lee}, {Line}, {Long},
  {Maire}, {Marchis}, {Matthews}, {Max}, {Metchev}, {Millar-Blanchaer},
  {Mittal}, {Morley}, {Morzinski}, {Murray-Clay}, {Oppenheimer}, {Palmer},
  {Patel}, {Perrin}, {Poyneer}, {Rafikov}, {Rantakyr\"{o}}, {Rice}, {Rojo},
  {Rudy}, {Ruffio}, {Ruiz}, {Sadakuni}, {Saddlemyer}, {Salama}, {Savransky},
  {Schneider}, {Sivaramakrishnan}, {Song}, {Soummer}, {Thomas}, {Vasisht},
  {Wallace}, {Ward-Duong}, {Wiktorowicz}, {Wolff}, \& {Zuckerman}}]{mgb+15}
{Macintosh}, B., {Graham}, J.~R., {Barman}, T., {et~al.} 2015, Sci, 350, 64+,
  \dodoi{10.1126/science.aac5891}

\bibitem[{{Maire} {et~al.}(2019){Maire}, {Rodet}, {Cantalloube}, {Galicher},
  {Brandner}, {Messina}, {Lazzoni}, {Mesa}, {Melnick}, {Carson}, {Samland},
  {Biller}, {Boccaletti}, {Wahhaj}, {Beust}, {Bonnefoy}, {Chauvin}, {Desidera},
  {Langlois}, {Henning}, {Janson}, {Olofsson}, {Rouan}, {M{\'e}nard},
  {Lagrange}, {Gratton}, {Vigan}, {Meyer}, {Cheetham}, {Beuzit}, {Dohlen},
  {Avenhaus}, {Bonavita}, {Claudi}, {Cudel}, {Daemgen}, {D'Orazi}, {Fontanive},
  {Hagelberg}, {Le Coroller}, {Perrot}, {Rickman}, {Schmidt}, {Sissa}, {Udry},
  {Zurlo}, {Abe}, {Orign{\'e}}, {Rigal}, {Rousset}, {Roux}, \&
  {Weber}}]{Maire2019}
{Maire}, A.~L., {Rodet}, L., {Cantalloube}, F., {et~al.} 2019, \aap, 624, A118,
  \dodoi{10.1051/0004-6361/201935031}

\bibitem[{{Marois} {et~al.}(2008){Marois}, {Macintosh}, {Barman}, {Zuckerman},
  {Song}, {Patience}, {Lafreni{\`e}re}, \& {Doyon}}]{Marois2008}
{Marois}, C., {Macintosh}, B., {Barman}, T., {et~al.} 2008, Science, 322, 1348,
  \dodoi{10.1126/science.1166585}

\bibitem[{{McLean} {et~al.}(2012){McLean}, {Berger}, \& {Reiners}}]{McLean2012}
{McLean}, M., {Berger}, E., \& {Reiners}, A. 2012, \apj, 746, 23,
  \dodoi{10.1088/0004-637X/746/1/23}

\bibitem[{{McMullin} {et~al.}(2007){McMullin}, {Waters}, {Schiebel}, {Young},
  \& {Golap}}]{McMullin2007}
{McMullin}, J.~P., {Waters}, B., {Schiebel}, D., {Young}, W., \& {Golap}, K.
  2007, in Astronomical Society of the Pacific Conference Series, Vol. 376,
  Astronomical Data Analysis Software and Systems XVI, ed. R.~A. {Shaw},
  F.~{Hill}, \& D.~J. {Bell}, 127

\bibitem[{{Naud} {et~al.}(2014){Naud}, {Artigau}, {Malo}, {Albert}, {Doyon},
  {Lafreni{\`e}re}, {Gagn{\'e}}, {Saumon}, {Morley}, {Allard}, {Homeier},
  {Beichman}, {Gelino}, \& {Boucher}}]{Naud2014}
{Naud}, M.-E., {Artigau}, {\'E}., {Malo}, L., {et~al.} 2014, \apj, 787, 5,
  \dodoi{10.1088/0004-637X/787/1/5}

\bibitem[{{Offringa} {et~al.}(2010){Offringa}, {de Bruyn}, {Biehl}, {Zaroubi},
  {Bernardi}, \& {Pandey}}]{odbb+10}
{Offringa}, A.~R., {de Bruyn}, A.~G., {Biehl}, M., {et~al.} 2010, MNRAS, 405,
  155, \dodoi{10.1111/j.1365-2966.2010.16471.x}

\bibitem[{{Offringa} {et~al.}(2012){Offringa}, {van de Gronde}, \&
  {Roerdink}}]{ovdgr12}
{Offringa}, A.~R., {van de Gronde}, J.~J., \& {Roerdink}, J. B. T.~M. 2012,
  A\&A, 539, A95+, \dodoi{10.1051/0004-6361/201118497}

\bibitem[{{Pineda} {et~al.}(2017){Pineda}, {Hallinan}, \& {Kao}}]{Pineda2017}
{Pineda}, J.~S., {Hallinan}, G., \& {Kao}, M.~M. 2017, \apj, 846, 75,
  \dodoi{10.3847/1538-4357/aa8596}

\bibitem[{{Rodriguez} {et~al.}(2011){Rodriguez}, {Zuckerman}, {Melis}, \&
  {Song}}]{Rodriguez2011}
{Rodriguez}, D.~R., {Zuckerman}, B., {Melis}, C., \& {Song}, I. 2011, \apjl,
  732, L29, \dodoi{10.1088/2041-8205/732/2/L29}

\bibitem[{{Rosenthal} {et~al.}(2021){Rosenthal}, {Fulton}, {Hirsch},
  {Isaacson}, {Howard}, {Dedrick}, {Sherstyuk}, {Blunt}, {Petigura}, {Knutson},
  {Behmard}, {Chontos}, {Crepp}, {Crossfield}, {Dalba}, {Fischer}, {Henry},
  {Kane}, {Kosiarek}, {Marcy}, {Rubenzahl}, {Weiss}, \&
  {Wright}}]{Rosenthal2021}
{Rosenthal}, L.~J., {Fulton}, B.~J., {Hirsch}, L.~A., {et~al.} 2021, arXiv
  e-prints, arXiv:2105.11583.
\newblock \doarXiv{2105.11583}

\bibitem[{{Route} \& {Wolszczan}(2012)}]{rw12}
{Route}, M., \& {Wolszczan}, A. 2012, ApJL, 747, L22+,
  \dodoi{10.1088/2041-8205/747/2/l22}

\bibitem[{{Route} \& {Wolszczan}(2013)}]{rw2013}
---. 2013, \apj, 773, 18, \dodoi{10.1088/0004-637X/773/1/18}

\bibitem[{{Route} \& {Wolszczan}(2016{\natexlab{a}})}]{rw16}
---. 2016{\natexlab{a}}, ApJL, 821, L21+, \dodoi{10.3847/2041-8205/821/2/L21}

\bibitem[{{Route} \& {Wolszczan}(2016{\natexlab{b}})}]{Route2016}
---. 2016{\natexlab{b}}, \apjl, 821, L21, \dodoi{10.3847/2041-8205/821/2/L21}

\bibitem[{{Sault} \& {Wieringa}(1994)}]{the.mfs}
{Sault}, R.~J., \& {Wieringa}, M.~H. 1994, A\&AS, 108, 585.
\newblock \url{http://adsabs.harvard.edu/abs/1994A\%26AS..108..585S}

\bibitem[{{Stassun} {et~al.}(2017){Stassun}, {Collins}, \&
  {Gaudi}}]{Stassun2017}
{Stassun}, K.~G., {Collins}, K.~A., \& {Gaudi}, B.~S. 2017, \aj, 153, 136,
  \dodoi{10.3847/1538-3881/aa5df3}

\bibitem[{{Stroe} {et~al.}(2012){Stroe}, {Snellen}, \&
  {R{\"o}ttgering}}]{Stroe2012}
{Stroe}, A., {Snellen}, I.~A.~G., \& {R{\"o}ttgering}, H.~J.~A. 2012, \aap,
  546, A116, \dodoi{10.1051/0004-6361/201220006}

\bibitem[{{Treumann}(2006)}]{t06}
{Treumann}, R. 2006, A\&ARv, 13, 229, \dodoi{10.1007/s00159-006-0001-y}

\bibitem[{{Turner} {et~al.}(2019){Turner}, {Grie{\ss}meier}, {Zarka}, \&
  {Vasylieva}}]{Turner2019}
{Turner}, J.~D., {Grie{\ss}meier}, J.-M., {Zarka}, P., \& {Vasylieva}, I. 2019,
  \aap, 624, A40, \dodoi{10.1051/0004-6361/201832848}

\bibitem[{{Turner} {et~al.}(2021){Turner}, {Zarka}, {Grie{\ss}meier}, {Lazio},
  {Cecconi}, {Emilio Enriquez}, {Girard}, {Jayawardhana}, {Lamy}, {Nichols}, \&
  {de Pater}}]{Turner2021}
{Turner}, J.~D., {Zarka}, P., {Grie{\ss}meier}, J.-M., {et~al.} 2021, \aap,
  645, A59, \dodoi{10.1051/0004-6361/201937201}

\bibitem[{{Vedantham} {et~al.}(2020){Vedantham}, {Callingham}, {Shimwell},
  {Tasse}, {Pope}, {Bedell}, {Snellen}, {Best}, {Hardcastle}, {Haverkorn},
  {Mechev}, {O'Sullivan}, {R{\"o}ttgering}, \& {White}}]{Vedantham2020}
{Vedantham}, H.~K., {Callingham}, J.~R., {Shimwell}, T.~W., {et~al.} 2020,
  Nature Astronomy, 4, 577, \dodoi{10.1038/s41550-020-1011-9}

\bibitem[{{White} {et~al.}(1989){White}, {Jackson}, \& {Kundu}}]{White1989}
{White}, S.~M., {Jackson}, P.~D., \& {Kundu}, M.~R. 1989, \apjs, 71, 895,
  \dodoi{10.1086/191401}

\bibitem[{{Williams}(2017)}]{w17}
{Williams}, P. K.~G. 2017, in Handbook of Exoplanets, ed. H.~J. {Deeg} \& J.~A.
  {Belmonte} (Springer Verlag).
\newblock \url{http://adsabs.harvard.edu/abs/2017arxiv170704264}

\bibitem[{{Williams} {et~al.}(2013){Williams}, {Berger}, \& {Zauderer}}]{wbz13}
{Williams}, P. K.~G., {Berger}, E., \& {Zauderer}, B.~A. 2013, ApJL, 767, L30+,
  \dodoi{10.1088/2041-8205/767/2/l30}

\bibitem[{{Williams} {et~al.}(2014){Williams}, {Cook}, \& {Berger}}]{wcb2014}
{Williams}, P.~K.~G., {Cook}, B.~A., \& {Berger}, E. 2014, \apj, 785, 9,
  \dodoi{10.1088/0004-637X/785/1/9}

\bibitem[{{Williams} {et~al.}(2017{\natexlab{a}}){Williams}, {Gizis}, \&
  {Berger}}]{wgb17}
{Williams}, P. K.~G., {Gizis}, J.~E., \& {Berger}, E. 2017{\natexlab{a}}, ApJ,
  834, 117+, \dodoi{10.3847/1538-4357/834/2/117}

\bibitem[{{Williams} {et~al.}(2017{\natexlab{b}}){Williams}, {Gizis}, \&
  {Berger}}]{Williams2017}
{Williams}, P.~K.~G., {Gizis}, J.~E., \& {Berger}, E. 2017{\natexlab{b}}, \apj,
  834, 117, \dodoi{10.3847/1538-4357/834/2/117}

\bibitem[{{Wu} \& {Lee}(1979)}]{Wu1979}
{Wu}, C.~S., \& {Lee}, L.~C. 1979, \apj, 230, 621, \dodoi{10.1086/157120}

\bibitem[{{Zarka}(1998)}]{zarka1998}
{Zarka}, P. 1998, \jgr, 103, 20159, \dodoi{10.1029/98JE01323}

\bibitem[{{Zarka} {et~al.}(2015){Zarka}, {Lazio}, \& {Hallinan}}]{Zarka2015}
{Zarka}, P., {Lazio}, J., \& {Hallinan}, G. 2015, in Advancing Astrophysics
  with the Square Kilometre Array (AASKA14), 120

\bibitem[{{Zurlo} {et~al.}(2016){Zurlo}, {Vigan}, {Galicher}, {Maire}, {Mesa},
  {Gratton}, {Chauvin}, {Kasper}, {Moutou}, {Bonnefoy}, {Desidera}, {Abe},
  {Apai}, {Baruffolo}, {Baudoz}, {Baudrand}, {Beuzit}, {Blancard},
  {Boccaletti}, {Cantalloube}, {Carle}, {Cascone}, {Charton}, {Claudi},
  {Costille}, {de Caprio}, {Dohlen}, {Dominik}, {Fantinel}, {Feautrier},
  {Feldt}, {Fusco}, {Gigan}, {Girard}, {Gisler}, {Gluck}, {Gry}, {Henning},
  {Hugot}, {Janson}, {Jaquet}, {Lagrange}, {Langlois}, {Llored}, {Madec},
  {Magnard}, {Martinez}, {Maurel}, {Mawet}, {Meyer}, {Milli},
  {Moeller-Nilsson}, {Mouillet}, {Orign{\'e}}, {Pavlov}, {Petit}, {Puget},
  {Quanz}, {Rabou}, {Ramos}, {Rousset}, {Roux}, {Salasnich}, {Salter},
  {Sauvage}, {Schmid}, {Soenke}, {Stadler}, {Suarez}, {Turatto}, {Udry},
  {Vakili}, {Wahhaj}, {Wildi}, \& {Antichi}}]{Zurlo2016}
{Zurlo}, A., {Vigan}, A., {Galicher}, R., {et~al.} 2016, \aap, 587, A57,
  \dodoi{10.1051/0004-6361/201526835}

\end{thebibliography}
\bibliographystyle{aasjournal}

\end{document}